\documentclass[12pt]{iopart}
\usepackage{citesort}
\usepackage{graphicx}
\usepackage{epstopdf}
\newcommand{\ket}[1]{{\left| {#1} \right\rangle}}

\newcommand{\ketbra}[2]{{\left| {#1} \right\rangle \!\!\left\langle {#2} \right|}}
\begin{document}

\title{Rotation sensing with trapped ions}

\author{W C Campbell$^{1,2}$ and P Hamilton$^1$}
\address{$^1$ UCLA Department of Physics and Astronomy, Los Angeles,
  California 90095, USA}
\ead{wes@physics.ucla.edu}
\address{$^2$ California Institute for Quantum Emulation, Santa
  Barbara, California 93106, USA}

\begin{abstract}
We present a protocol for using trapped ions to measure rotations via
matter-wave Sagnac interferometry.  The trap allows the interferometer
to enclose a large area in a compact apparatus through repeated
round-trips in a Sagnac geometry.  We show how a uniform magnetic
field can be used to close the interferometer over a large dynamic
range in rotation speed and measurement bandwidth without losing
contrast.  Since this technique does not require the ions to be
confined in the Lamb-Dicke regime, thermal states with many phonons
should be sufficient for operation.

\end{abstract}
\maketitle

\section{Introduction}
The Sagnac effect can be used to measure the rotational velocity
\boldmath$\Omega$ \unboldmath of a reference frame by observing the
phase shift of an interferometer in that frame whose paths enclose an
area $A$ perpendicular to any component of
\boldmath$\Omega$ \unboldmath (see, \textit{e.g.} \cite{BarrettCRP14}
for a review). The rotation-induced phase shift is given by
\begin{equation}
\Phi=2 \pi\frac{2 E}{h c^2}\mathbf{A}\cdot
\mbox{\boldmath$\Omega$\unboldmath} ,\label{SagnacPhase}
\end{equation}
where $\mathbf{A}$ is the vector area enclosed by the two paths. $E$
is the total energy of the particles that are interfering, defined
using the relativistic energy-momentum relation
\begin{equation}
E^2 = \left(mc^2\right)^2 + p^2c^2.
\end{equation}
For photons, $E=\hbar \omega_{\mathrm{optical}}$, whereas for atoms of
rest mass $m$ moving at non-relativistic speeds ($p\ll mc$), $E=mc^2$.

The sensitivity of a gyroscope is the minimum detectable rotation rate
within a detection bandwidth $\Delta f$.  For a shot-noise-limited
interferometer that detects the outcome of individual interference
events at a rate $\dot{N}$, the uncertainty in the measured phase
after running for a time $t = 1/\Delta f$ will be $\delta \phi \approx
\sqrt{\Delta f/\dot{N}}$.  The sensitivity is given by
\begin{equation}
\mathcal{S} = \frac{\delta \phi}{\frac{\partial \Phi}{\partial \Omega}\sqrt{\Delta
    f}}=\frac{1}{\frac{\partial \Phi}{\partial \Omega}\sqrt{\dot{N}}}
\end{equation}
where the \textit{scale factor} is given by
\begin{equation}
\frac{\partial \Phi}{\partial \Omega} = 2 \pi A \frac{2E}{hc^2} \label{ScaleFactor}
\end{equation}
and we have assumed an orientation such
that $\mathbf{A} \cdot$\boldmath$\Omega$\unboldmath$= A\Omega$ for
algebraic simplicity.

Two primary methods are frequently employed to boost the sensitivity
of interferometric gyroscopes.  For photons, optical fibers (or ring
laser cavities) allow many effective round-trips through the Sagnac
interferometer, thereby increasing the effective area $A$ by 2 times
the number of round trips ($M$) without increasing the actual area of
the apparatus.  This, coupled with the large $\dot{N}$ possible
in these devices, leads to a state-of-the-art reported sensitivity of
$\mathcal{S} = 1.2 \times 10^{-11} \mbox{
  rad}/\mbox{s}/\sqrt{\mbox{Hz}}$, achieved by a ring laser with $16
\mbox{ m}^2$ enclosed area \cite{SchreiberRSI13}.

Another approach is to use atoms on ballistic trajectories instead of
photons, which increases $E$ by a factor of $mc^2/\hbar
\omega_{\mathrm{optical}} \approx 10^{11}$.  The drawbacks of this
approach, as compared to an optical gyroscope, are that
$\dot{N}$ is smaller, and the free-flight atom trajectories
enclose the interferometer area $A$ only once.  The latter constraint
has meant that increasing $A$ has necessarily involved increasing the
physical size of the apparatus, which can be undesirable for some
applications.  Furthermore, long atom trajectories and large
separations make the measurement susceptible to systematics that can
produce path-dependent phase shifts, such as magnetic field gradients.
Nonetheless, the improvement in $E$ has enabled atom interferometers
to demonstrate high rotation rate sensitivity, with demonstrated
state-of-the-art short-term sensitivities of $\mathcal{S} = 6 \times
10^{-10} \mbox{ rad}/\mbox{s}/\sqrt{\mbox{Hz}}$ for atomic beams
\cite{GustavsonCQG00} and $\mathcal{S} = 2.4 \times 10^{-7} \mbox{
  rad}/\mbox{s}/\sqrt{\mbox{Hz}}$ for laser-cooled atoms
\cite{GauguetPRA09}.

Here, we show that trapped atomic ions provide a way to use both
methods simultaneously to increase the interferometer sensitivity.
While interferometers with enclosed area have been demonstrated with
clouds of trapped neutral atoms \cite{SackettPRA09,PrentissPRL07},
maintaining the coherence across the ensemble needed for a gyroscope
has proved difficult.  Here we introduce a combination of laser-driven
spin-dependent momentum kicks in one direction with ion trap voltage
changes along an orthogonal direction that perform interferometry with
trapped ions in a Sagnac (as opposed to Mach-Zehnder) configuration.
This allows atomic trajectories to repeatedly enclose the same area,
thereby accumulating Sagnac phase continuously for a time that is not
limited by a ballistic flight trajectory.  Since the enclosed area is
proportional to the displacement along both directions and only one of
these needs to be state-dependent, the interferometer area can be
increased with trap voltage alone, circumventing the need to drive
more coherent momentum transfer from the laser.  The harmonic trapping
potential makes the area enclosed independent of the initial ion
velocity, eliminating a source of scale factor instability found in
free space atom interferometers.  These factors, coupled with the
extremely long coherence times of trapped ions, gives the trapped ion
interferometer the potential to enclose a large effective area in a
small apparatus with high stability.

\begin{figure}
\begin{center}
\includegraphics[scale=0.5]{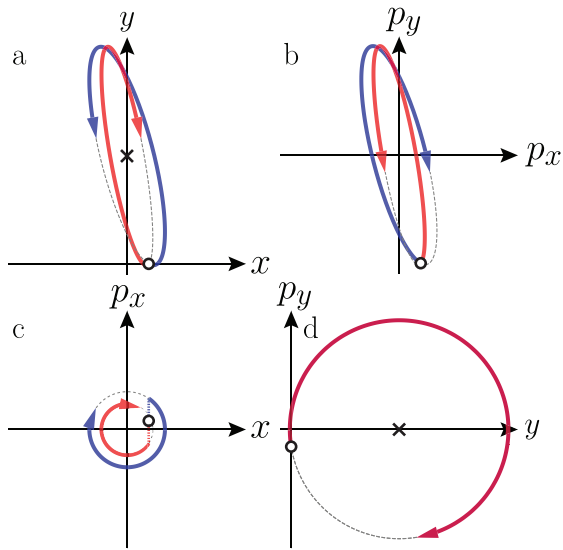}
\end{center}
\caption{Trajectories of an ion during interferometer operation in
  (a) position space, (b) momentum space (c) $x$ phase space and (d)
  $y$ phase space.  The ion's starting coordinates are indicated by
  a circle, and the trap center after the $y$-displacement in
  step (\ref{Stepiii}) is indicated by an $\times$.  Red and blue
  curves represent the trajectory for the two spin states.
  Trajectories for different starting conditions are qualitatively
  similar to these, with the exception that for a ground-state ion,
  the trajectories for the two spin states completely overlap.}
\label{TrajectoriesFigure}
\end{figure}

\section{Interferometer operation}
We begin by introducing the protocol for measuring rotations with a
single ion that hosts a qubit with internal states $\ket{\uparrow}$
and $\ket{\downarrow}$.  As shown in Fig.~\ref{TrajectoriesFigure},
the enclosed area will be in the $x,y$ plane, and the (secular) trap
frequencies for the ion in these two directions after the $y$
displacement (see below) will be assumed to be degenerate: $\omega_x =
\omega_y \equiv \omega$.  $(x,y,z)$ will be coordinates in real space,
while $(X,Y,Z)$ will denote axes of the qubit's associated Bloch
sphere.  We will assume that the confinement in the $z$-direction is
strong ($\omega_z \gg \omega$) so that the system can be approximated
as being 2D.  The time-sequence of the trapped ion gyroscope proceeds
in the following steps (see also Fig.~\ref{TrajectoriesFigure} and
\ref{RotatingFrameFigure}):
\begin{enumerate}
\item Prepare the ion in $\ket{\downarrow}$ and apply a $\pi/2$ pulse
  of microwaves about $-\mathbf{\hat{Y}}.$\label{Stepi}
\item Apply $N_{\mathrm{k}}$ spin-dependent kicks (SDKs) in the
  $x$-direction ($\Delta \mathbf{p} = -N_{\mathrm{k}}\hbar \Delta
  \mathbf{k}\hat{\sigma}_Z $) to separate the atom in momentum
  space.\label{Stepii}
\item Apply a step function in electrode voltages to non-adiabatically
  displace the trap only in the $y$-direction a distance $y_{\mathrm{d}}$.\label{Stepiii}
\item Allow the ion to oscillate in the trap for an integer number ($M$) of
  round trips $\Delta t = M 2\pi/\omega$.\label{Stepiv}
\item Reverse step (\ref{Stepiii}) by non-adiabatically switching the
  trap voltages back to their original values.\label{Stepv}
\item Reverse step (\ref{Stepii}) by applying $N_{\mathrm{k}}$ SDKs
  in the other direction ($\Delta \mathbf{p} = N_{\mathrm{k}}\hbar \Delta
  \mathbf{k}\hat{\sigma}_Z $) to close the interferometer.\label{Stepvi}
\item Apply another $\pi/2$ pulse with microwaves about an
  axis inclined by $\phi$ in the $X,Y$ plane from the $-\mathbf{\hat{Y}}$ axis of the
  Bloch sphere, then measure the internal state of the ion in the
  qubit basis.\label{Stepvii}
\end{enumerate}

We note that steps (\ref{Stepi}) and (\ref{Stepvii}) are a standard
Ramsey sequence, so qubit and microwave oscillator coherence are
required for the duration of the protocol. Even for non-clock-state
qubits with magnetic sensitivity on the order of a Bohr magneton
($\mu_{\mathrm{B}}$), qubit coherence times of the order $1 \mbox{ s}$
or greater can be achieved \cite{RusterArXiv}.  We will
describe the details of the gyroscope protocol assuming the magnetic
field on the ion is zero before discussing the magnetic field effects.

\section{Phase-space displacements}
The trapped ion gyroscope relies on two different methods to produce
displacements in motional phase space: spin-dependent kicks that
transfer photon momenta to the ions in directions that depend upon the
ion's spin (qubit) state, and trap voltage steps that rapidly displace
the trap center. Since a displacement operation in phase space
necessarily involves a large number of Fock states, both of these
operations take place much faster than the resolved-sideband limit ($T
\approx 2 \pi/ \omega$), and can be thought of as driving many
different motional transitions at once.

\begin{figure}
\begin{center}
\includegraphics[scale=1]{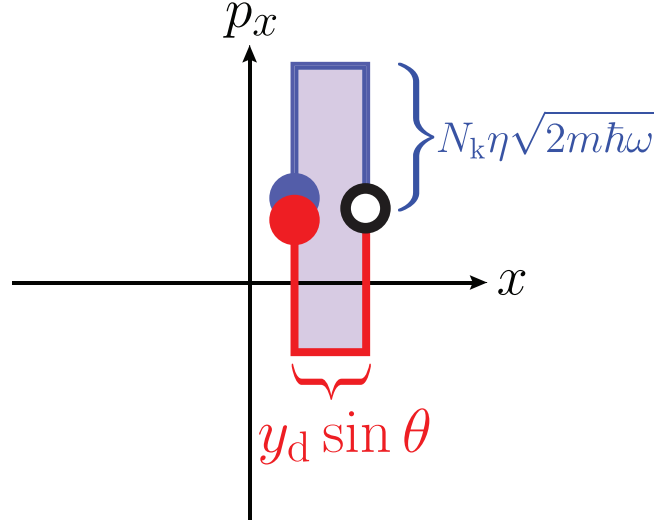}
\end{center}
\caption{Trajectory of an ion in $x$ phase space in the interaction
  picture with respect to the harmonic oscillation.  The ion's
  starting coordinates are indicated by a circle, and red and blue
  curves represent the trajectory for the two spin states.  A
  freely-evolving coherent state in this ``rotating frame'' (rotating
  in phase space, as opposed to real space) appears stationary; the
  trajectories shown are induced by the displacement operators.  For
  small rotations ($\theta \ll 1$), the area enclosed in this phase
  space is the Sagnac phase (\ref{SagnacExpanded}) for an ion that
  starts at position $y=0$.}
\label{RotatingFrameFigure}
\end{figure}

The spin-dependent kicks \cite{MizrahiPRL13} of steps (\ref{Stepii})
and (\ref{Stepvi}) act as the beam splitters in the matter-wave
interferometer.  The speed of the SDK is enabled through the
utilization of ``ultrafast'' mode-locked lasers to transfer $\hbar
\Delta k$ of momentum to the ion (see \cite{MizrahiPRL13,MizrahiAPB14} for
more detail).  Conceptually, the ideal SDK transfers a momentum kick
to the ion whose direction is reversed for the two spin states via the
operator
\begin{equation}
\hat{U}_{\mathrm{SDK}} = \hat{D}_x[\rmi \eta] \hat{\sigma}_+ +
\hat{D}_x[-\rmi \eta] \hat{\sigma}_-\label{SDKOperator}
\end{equation}
where $\hat{D}_x[s]$ displaces a coherent state in $x$ phase space a
distance $s$ (see Fig.~\ref{TrajectoriesFigure} and
\ref{RotatingFrameFigure}) and $\eta$ is the Lamb-Dicke factor for the
laser-ion interaction in the $x$-direction ($\eta \equiv \Delta k x_0
= \Delta k \sqrt{\hbar/2 m \omega}$).  Since step (\ref{Stepvi})
drives the same process as (\ref{Stepii}) with the direction of the
kicks reversed (effectively replacing $\rmi \!\rightarrow -\rmi$ in
(\ref{SDKOperator})), we have suppressed the laser beat note phase
when writing (\ref{SDKOperator}) since it plays no role as along as it
is stable during a single enactment the interferometer protocol.  The
qubit raising and lowering operators ($\hat{\sigma}_{\pm}$) flip the
spin state of the qubit, so one way that larger displacements
(\textit{i.e.} $M$ of them) can be made is by repeating this operation
after a delay by half a motional period \cite{JohnsonPRL15}.  For
algebraic simplicity, we will assume for our protocol that the number
of spin-dependent kicks applied ($N_{\mathrm{k}}$) is even and that an
extra half-period of motion is inserted after the last kick to
preserve the harmonic oscillation phase of the initial motional state.

Working in the coherent state basis for describing the ion's motion in
$x$ and $y$ (denoted by coherent state parameters $\alpha_x$ and
$\alpha_y$), step (\ref{Stepi}) results in the state
\begin{equation}
\ket{\psi_{\mathrm{\ref{Stepi}}}} = \textstyle \frac{1}{\sqrt{2}} \displaystyle
\left( \ket{\downarrow} + \ket{\uparrow} \right) \otimes \ket{\alpha_x,\alpha_y}.
\end{equation}
The SDKs in step (\ref{Stepii}) induce spin-orbit coupling to produce the
entangled state
\begin{eqnarray}
\fl \ket{\psi_{\mathrm{\ref{Stepii}}}} = &\textstyle \frac{1}{\sqrt{2}} \displaystyle
\big( \rme^{\rmi N_{\mathrm{k}} \eta \mathrm{I\!R}(\alpha_x)}
\ket{\downarrow} \otimes \ket{\alpha_x + \rmi N_{\mathrm{k}}\eta}
\nonumber \\ 
& + \rme^{-\rmi N_{\mathrm{k}} \eta \mathrm{I\!R}(\alpha_x)}
\ket{\uparrow} \otimes \ket{\alpha_x - \rmi N_{\mathrm{k}}\eta} \big)
\otimes \ket{\alpha_y}
\end{eqnarray}
where $\mathrm{I\!R}(\alpha)$ denotes the real part of the coherent
state parameter $\alpha$.

Interferometers based on SDKs have been proposed \cite{PoyatosPRA96}
to measure the Sagnac effect, and have recently been implemented in a
1D non-Sagnac geometry to measure temperature over a wide dynamic
range \cite{JohnsonPRL15}. However, for a Sagnac gyroscope, the second
displacement need not be spin-dependent and can therefore be implemented as a
simple trap center shift.  Electrode voltages can be rapidly changed
to displace the trap center, an operation that has been demonstrated
to couple to thousands of Fock states to perform a coherent state
displacement operation \cite{AlonsoNatComms16}.  Compared to coherent
momentum transfer from a laser, voltage-driven motion of this sort can
produce a larger displacement, and does so without the detrimental
effects of spontaneous emission and differential AC Stark shifts
associated with laser-driven gates.

A rapid shift of the trap center by a physical distance
$y_{\mathrm{d}}$ along $y$ can be modeled with a displacement operator
\begin{equation}
\hat{D}_y\!\big[ \textstyle
  -\frac{y_{\mathrm{d}}}{2 y_0} \displaystyle \big] \ket{\alpha_y} =
\big| \alpha_y - \textstyle
  \frac{y_{\mathrm{d}}}{2 y_0} \displaystyle \big\rangle
\end{equation}
where $y_0 = x_0 \equiv \sqrt{\hbar/2 m \omega}$ and we will refrain
from writing phase terms that for this protocol are global.

\section{Rotation-induced phase}
The effect of rotation in this system can be described in either the
non-rotating frame (the ion's frame) or the rotating reference frame
(the apparatus frame).  We choose the former, which means that the
rotation manifests itself as change in the direction of the kicks in
steps (\ref{Stepv}) and (\ref{Stepvi}).  A constant rotation rate
$\Omega$ about the positive $z$-axis of the apparatus will shift the
angles of these kicks by
\begin{equation}
\theta = \Omega \Delta t = \Omega M \frac{2 \pi}{\omega}.\label{thetaDef}
\end{equation}
This transforms the displacement operators according to
\begin{equation}
\hat{D}^{\prime} = \rme^{-\rmi \theta \hat{J}_z}\hat{D} \,\rme^{\rmi \theta \hat{J}_z}
\end{equation}
and the state
of the ion after step (\ref{Stepvi}) is

\begin{eqnarray}
\fl \ket{\psi_{\mathrm{\ref{Stepvi}}}} = &\textstyle \frac{1}{\sqrt{2}} \displaystyle
\Big( \rme^{\rmi \delta/2}
\ket{\downarrow} \otimes \big| \alpha_x + \rmi N_{\mathrm{k}}\eta(1 - \cos
  \theta) - \textstyle
  \frac{y_{\mathrm{d}}}{2 x_0} \displaystyle \sin \theta \big\rangle
\nonumber \\ 
& \;\; \otimes \big| \alpha_y - \textstyle
  \frac{y_{\mathrm{d}}}{2 y_0} \displaystyle(1 - \cos
  \theta) - \rmi N_{\mathrm{k}} \eta \sin \theta \big\rangle \nonumber \\
& + \rme^{-\rmi \delta/2}
\ket{\uparrow} \otimes \big| \alpha_x - \rmi N_{\mathrm{k}}\eta(1 - \cos
  \theta) - \textstyle
  \frac{y_{\mathrm{d}}}{2 x_0} \displaystyle \sin \theta \big\rangle
  \nonumber \\
& \;\; \otimes \big| \alpha_y - \textstyle
  \frac{y_{\mathrm{d}}}{2 y_0} \displaystyle(1 - \cos
  \theta) + \rmi N_{\mathrm{k}} \eta \sin \theta \big\rangle \Big)\label{psivi}
\end{eqnarray}
where the relative phase ($\delta$) is given by
\begin{eqnarray}
\delta &=& 2 N_{\mathrm{k}} \eta \Big( \textstyle \frac{y_{\mathrm{d}}}{2
  x_0} \displaystyle (1 + \cos \theta )\sin \theta + \textstyle \frac{y_{\mathrm{d}}}{2
  y_0} \displaystyle (1 - \cos \theta )\sin \theta    \nonumber \\
 && \;\;\;\;\;\;\;+ \mathrm{I\!R}(\alpha_x)( 1 - \cos \theta) -
\mathrm{I\!R}(\alpha_y)\sin \theta \Big).
\end{eqnarray}

(\ref{psivi}) shows that this protocol leaves residual entanglement
between the spin and motion in both $x$ and $y$.  Using
$\ket{\mu_{i,\mathrm{f}}(\theta)}$ to denote the final motional states
in $x$ and $y$ for the parts of the wavefunction that are associated
with spin state $i \in \{ \uparrow,\downarrow\}$ in (\ref{psivi}),
the overlap is
\begin{equation}
\big\langle \mu_{\downarrow,\mathrm{f}}(\theta) \big|
\mu_{\uparrow,\mathrm{f}}(\theta) \big\rangle =
\rme^{-2(2N_{\mathrm{k}}\eta \sin\frac{\theta}{2})^2} \rme^{-\rmi \delta^{\prime}}
\end{equation}
where the first term comes from the imperfect state overlap (which is
confined entirely to momentum space) and the second is a pure phase
term called the \textit{overlap phase} $\delta^{\prime}$:
\begin{equation}
\delta^{\prime} \equiv 2 N_{\mathrm{k}}\eta \big(
\mathrm{I\!R}(\alpha_x)(1 - \cos \theta) - \mathrm{I\!R}(\alpha_y)\sin
\theta \big).
\end{equation}
Residual entanglement between spin and motion will reduce the contrast
of the interferometer, and it is the sum of $\delta$ and $\delta^{\prime}$
that contributes the phase shift that is measured using this protocol.
However, as we show below, the phase shift that is
measured is unaffected by the initial ion temperature, and there is no
requirement that this device be operated in the Lamb-Dicke regime.

\section{Readout}
In order to measure the rotation-induced phase ($\delta +
\delta^{\prime}$), step (\ref{Stepvii}) applies a second Ramsey zone
with a controllable phase shift $\phi$, yielding
\begin{eqnarray}
\fl \ket{\psi_{\mathrm{\ref{Stepvii}}}} = &\textstyle \frac{1}{2}
\displaystyle \Big( \rme^{\rmi \delta/2} \left( \rme^{-\rmi \phi}
\ket{\uparrow} + \ket{\downarrow} \right) \otimes
\ket{\mu_{\downarrow,\mathrm{f}}(\theta)} \nonumber \\
& + \rme^{-\rmi
  \delta/2} \left( \ket{\uparrow} - \rme^{\rmi \phi} \ket{\downarrow}
\right) \otimes \ket{\mu_{\uparrow,\mathrm{f}}(\theta)}
\Big). \label{psivii}
\end{eqnarray}
This step maps the motional phase onto the internal state of the ion,
which would then measured using standard fluorescence techniques.  The
probability of measuring, for instance, spin up $(\ket{\uparrow})$ is
given by
\begin{equation}
\mathcal{P}(\uparrow, \theta, \phi) =  \!\!\int \!\! \mathrm{d}^2
\alpha_x \mathrm{d}^2 \alpha_y\,
P(\alpha_x) P(\alpha_y) \langle \psi_{\mathrm{\ref{Stepvii}}}
\ketbra{\uparrow}{\uparrow} \psi_{\mathrm{\ref{Stepvii}}} \rangle\label{MotionTrace}
\end{equation}
where 
\begin{eqnarray}
\fl \langle \psi_{\mathrm{\ref{Stepvii}}}
\ketbra{\uparrow}{\uparrow} \psi_{\mathrm{\ref{Stepvii}}} \rangle
= \frac{1}{2} + \frac{1}{2} \rme^{-2(2N_{\mathrm{k}}\eta
  \sin\frac{\theta}{2})^2} \nonumber \\
\times \cos \left(\!\phi -
\frac{A(\alpha_y)}{\pi x_0^2} \sin \theta - 4 N_{\mathrm{k}}\eta x_0^2
\mathrm{I\!R}(\alpha_x)(1\! - \!\cos \theta )\!
\right)\label{Preadout}
\end{eqnarray}
and $P(\alpha_j)$ is the Glauber-Sudarshan $P$-representation describing the
(potentially mixed) initial motional state in the coherent state
basis.  $A(\alpha_y)$ is the classical, geometric area of the ellipse
enclosed by the ion trajectories in the $x,y$ plane, which depends upon
the initial position in $y$ via $y_i = 2 y_0 \mathrm{I\!R}(\alpha_y)$:
\begin{equation}
A(\alpha_y) \equiv \pi \,2 x_0 N_{\mathrm{k}}\eta \, (y_{\mathrm{d}} -
2 y_0\mathrm{I\!R}(\alpha_y)). \label{Area}
\end{equation}
In \ref{AreaAppendix}, we show that a semiclassical derivation agrees
with this quantum calculation. If we expand the argument of the cosine
in (\ref{Preadout}) to first order in $\theta$, we see that it
simplifies to $\phi - \Phi$, where
\begin{equation}
\Phi = \frac{A(\alpha_y)}{\pi x_0^2}\theta = 2 \pi \frac{2 m c^2}{h
  c^2} (2 M A(\alpha_y)) \Omega. \label{SagnacExpanded}
\end{equation} 
This is identifiable as the Sagnac phase shift (\ref{SagnacPhase})
with an effective area of $A_{\mathrm{eff}} = 2 M A(\alpha_y)$ since
the ion encloses the ellipse area $A(\alpha_y)$ twice each period for
$M$ periods.  (\ref{SagnacExpanded}) also provides some insight into
the origin of the scale factor of this interferometer: the rotation
angle $\theta$ is effectively ``amplified'' by a gain factor of
$A(\alpha_y)/\pi x_0^2$, the ratio of the enclosed area to the area of
the ground-state wavefunction.  This gain factor is the angular
momentum of the ion's motion divided by $\hbar$, and the
interferometer can therefore be thought of as a generalized atomic or
nuclear spin gyroscope with a very large effective spin.

\section{Finite temperature}
For an ion that is pre-cooled to the motional ground state along $y$
($\alpha_y=0$), (\ref{SagnacExpanded}) gives precisely the desired
outcome (\ref{SagnacPhase}) for the trapped ion gyroscope.  For an ion
that is initially in a thermal state with mean phonon occupation numbers
$\bar{n}_x = \bar{n}_y\equiv \bar{n}$, (\ref{MotionTrace}) can be used
to calculate the probability of measuring spin up:
\begin{eqnarray}
\mathcal{P}(\uparrow, \theta, \phi)& =& \frac{1}{2} + \frac{1}{2} \rme^{-(4N_{\mathrm{k}}\eta
  \sin\frac{\theta}{2})^2(\bar{n} + \frac{1}{2})} \nonumber \\
&&\times \cos \left(\phi - \frac{A(0)}{\pi x_0^2}\sin \theta   \right), \label{ThermalStateSpinUp}
\end{eqnarray}
which is valid to all orders in $\theta$.  The effect of finite
temperature is a reduction in the contrast of the interference, but
does not produce a phase shift of the signal.  However, since the
exponent in (\ref{ThermalStateSpinUp}) is proportional to
$\sin^2(\theta/2)$ and the Sagnac phase shift is proportional to
$\sin(\theta)$, the free evolution time ($\Delta t$ in
(\ref{thetaDef})) can be chosen to satisfy
\begin{equation}
\sin^2\left( \frac{\theta}{2} \right) \ll 16 N_{\mathrm{k}}^2 \eta^2
\left( \bar{n} + \frac{1}{2} \right)
\end{equation}
and the interferometer can be operated at essentially full contrast,
even at high temperature.  There is therefore no requirement that the
ion be cooled to the Lamb-Dicke regime, and as we estimate below,
Doppler cooling should be sufficient for full-contrast operation.

\section{Magnetic field effects}
Since the ion is moving while it is accumulating rotation-induced
phase, a nonzero magnetic field will give rise to a Lorentz force on
the moving monopole.  For a magnetic field in the $z$-direction, this
will cause the ion's orbit to precess in the $x,y$ plane, which will
lead to a false rotation signal.  Specifically, the
magnetically-induced rotation rate ($\Omega_{\mathrm{m}}$) can be
found \cite{SakuraiPRD80} by equating the Lorentz and Coriolis forces
for a uniform, static magnetic field in the $z$-direction ($\mathbf{B}
= B_z \mathbf{\hat{z}}$):
\begin{equation}
2 m \Omega_{\mathrm{m}} ( \mathbf{v} \times \mathbf{\hat{z}}) =
e B_z (\mathbf{v} \times \mathbf{\hat{z}}).
\end{equation}
This rotation rate is half the cyclotron frequency
$\omega_{\mathrm{cyc}}= e B/m = 2 \Omega_{\mathrm{m}}$, and the
precession angle this will produce is boosted up by the gain factor of
$A(0)/\pi x_0^2$ to give a magnetically-induced phase shift of
\begin{equation}
\Phi_{\mathrm{m}} \equiv \Delta t\, \Omega_{\mathrm{m}}
\frac{A(0)}{\pi x_0^2} =
\Delta t \frac{2}{\hbar} \left(e \frac{\omega}{2 \pi} \right) A(0) B_z.
\end{equation}
This phase shift can be interpreted as the dynamical phase from the
Zeeman shift of the ion's motional magnetic moment,
\begin{equation}
\mu_{\mathrm{m}} \equiv \frac{1}{2}\frac{\partial \left( \hbar
  \frac{\Phi_{\mathrm{m}}}{\Delta t} \right)}{\partial B_z} = I A(0)
\end{equation}
where $I\equiv e\omega/2 \pi$ is the current from the ion's motion.
This matches the classical expression for the magnetic moment of a
current loop of area $A(0)$.

\section{Non-harmonic corrections}
The analysis we have presented has thus far assumed a perfectly
harmonic potential.  We can find the first-order phase correction for
small non-harmonic terms of the potential by treating these terms as a
perturbation and integrating over the unperturbed trajectories.
Assuming the potential remains separable the formulas below hold for
each axis.  Let us write the general potential as
\begin{equation}
V=\frac{1}{2}m\omega^2 x_l^2\left(\frac{x^2}{x_l^2}+C_3
\frac{x^3}{x_l^3}+ C_4 \frac{x^4}{x_l^4}+\cdots\right)
\end{equation}
where $x_l$ is a length scale for the amplitude of the ion's motion and the
$\{C_i\}$ are dimensionless numbers assumed to be much smaller than one.  With a
harmonic trajectory $x(t)=x_l \sin(\omega t+\phi)$, only even $i$ terms
are non-zero.  Integrating over $M$ orbits gives
\begin{eqnarray}
\Delta\phi&=\frac{1}{\hbar}\int_0^{2\pi M/\omega}\mathrm{d}t\, \frac{1}{2}m\omega^2 x_l^2\sum_{i\geq 3}{C_i\sin^i(\omega t+\phi)}\\
&=m\omega x_l^2\frac{3\pi M}{8\hbar}C_4+\cdots 
\end{eqnarray}

\section{Performance}
Once the evolution time has been fixed, the sensitivity
of the trapped ion gyroscope can be written
\begin{equation}
\mathcal{S} = \frac{1}{2 N_{\mathrm{k}} \Delta k \, y_{\mathrm{d}}\sqrt{\Delta t}}.
\end{equation}

Since this is independent of the trap frequency $\omega$ (we assume
$M\! \gg\! 1$ and can therefore be chosen essentially arbitrarily),
the trapped ion gyroscope can be operated in a relatively
low-frequency trap as compared to typical traps for applications
requiring resolved sideband operations.  This provides the practical
advantage of making the non-adiabatic operations easier to achieve
with high fidelity in a fixed time.  It also permits the use of a trap
whose electrodes are far apart and far from the ion, which will
suppress surface-induced heating and patch charge perturbations and
improve harmonicity for a fixed absolute length scale.

We also note that the performance of this rotation sensor is
independent of the mass of the ion, and depends essentially only on
the wavelength of the laser used to drive the SDKs.  We will estimate
parameters for ${}^{171}\mathrm{Yb}^+$, which was used for the first
demonstrations of spin-dependent kicks \cite{MizrahiPRL13}, but
estimates for other species will be similar in magnitude.

For the hyperfine clock-state qubit in ${}^{171}\mathrm{Yb}^+$,
stimulated Raman transitions can be driven by a tripled vanadate laser
at $4 \pi/\Delta k= 355 \mbox{ nm}$ \cite{CampbellPRL10} with
$N_{\mathrm{k}} = 100$ \cite{MonroePrivateCommunication}.  Since this
qubit has a demonstrated coherence time exceeding $1000 \mbox{ s}$
\cite{FiskIEEE97}, a free-evolution time of $\Delta t = 1\mbox{ s}$
should be straightforward to achieve.  For a (secular) trap frequency
of $\omega/2 \pi = 10 \mbox{ kHz}$, a trap displacement of
$y_{\mathrm{d}} = 100 \mbox{ }\mu\mbox{m}$ would correspond to
$\langle n \rangle \approx 9 \times 10^5$ phonons, where displacements
corresponding to $\langle n \rangle \approx 10^4$ have already been
demonstrated \cite{AlonsoNatComms16}.  Cooling $\mathrm{Yb}^+$ to the
Doppler limit ($T_{\mathrm{D}} = \hbar \gamma/ 2 k_{\mathrm{B}}$) in
such a trap will result in an interference contrast of 85\% for the
(sidereal) rotation rate of the earth $\Omega_{\mathrm{e}} \approx 73
\mbox{ }\mu\mbox{rad}/\mbox{s}$.  A trapped ion gyroscope operated
with these parameters would have a scale factor of $\partial \Phi /
\partial \Omega = 52 \mbox{ rad/}\Omega_{\mathrm{e}}$ and a
sensitivity of $\mathcal{S} = 1.4 \times 10^{-6} \mbox{
  rad/s/}\sqrt{\mbox{Hz}}$.  Improvements in the numbers of SDKs or
the distance of coherent trap displacements would make this
competitive with cold atom interferometers that use large numbers of
atoms.

For an interferometer using the parameters discussed above, the
magnetically-induced rotation rate per unit field is
$\Omega_{\mathrm{m}}/ (2 \pi B_z)= 4.5 \mbox{ Hz}/\mbox{G}$.  The
associated motional magnetic moment is $\mu_{\mathrm{m}} = 1.1
\mu_{\mathrm{B}}$, where $\mu_{\mathrm{B}}$ is the Bohr magneton.
Since the magnetic field stabilization required to combat this
systematic only needs to be applied to a small volume ($\ll 1 \mbox{
  cm}^3$), this magnetic sensitivity resembles the effect of using a
Zeeman-sensitive qubit, and many of the technical difficulties
associated with this have been overcome in various trapped ion quantum
information processing experiments \cite{RusterArXiv}.

In the limit where the magnetically-induced rotation rate is much
slower than the (secular) trap frequency ($\Omega_{\mathrm{m}} \ll
\omega$), the ion's motion is in elliptical orbits of fixed area whose
orientation slowly rotates.  Too much rotation will reduce the
contrast of the interference signal since the kicks in step
(\ref{Stepv}) and (\ref{Stepvi}) will not efficiently close the
interferometer loop.  However, trapped ions have also been
demonstrated as superb magnetic field sensors, and it seems likely
that with a periodic measurement of a stationary ion's Zeeman
splitting, a well-controlled field could be applied to cancel this
effect.

Likewise, magnetic rotation could be leveraged to cancel the contrast
reduction associated with high actual rotation rates (the exponential
factor in (\ref{Preadout})).  In this ``closed-loop mode,'' the
magnetic field needed to cancel the rotation would become the output
signal for the interferometer, and low-resolution rotation sensors
could be incorporated to feed forward the magnetic field needed to
keep the interferometer contrast maximized and on the steepest part of
a fringe.

\section{Discussion}
As compared to free-flight matter-wave interferometers, the trapped ion device provides many practical advantages.  First, the physical size of the interferometer can be compact while still retaining a large effective interferometer area by using multiple orbits.  Second,
since the ion wavepacket re-combines in space twice per trap period,
this interferometer can be interrogated over a wide dynamic range of
free-evolution times.  Fast rotation rates, which can be problematic
in neutral atom systems if the wavepackets don't re-combine or leave
the interferometry region, can be
compensated by applying uniform magnetic fields.  The operational
mode could be to actively stabilize the fringes with an applied field,
which becomes the readout signal.  There is also no need to keep
multiple optical beam paths interferometrically (relatively) stable
since the only steps that are sensitive to a laser phase (the SDKs,
steps (\ref{Stepii}) and (\ref{Stepvi})) are driven by the same laser
with its beam traversing the same optical path.  In addition, by using single ion wavepackets
which travel the same average trajectory, only in opposite directions, we eliminate spatially varying systematics.  Finally, free-flight interferometers have sensitivities to accelerations and the atomic beam velocities, whereas the scale factor for the ion trap interferometer depends only on the momentum kicks and trap displacement.

Another advantage of using trapped ions instead of neutral atoms for
matter-wave interferometry is the potential to leverage the advances
in trapped ion quantum information processing to produce
sub-shot-noise scaling of the sensitivity with ion number.  For
example, a collection of $N_{\mathrm{I}}$ ions could be prepared in
step (\ref{Stepi}) in a GHZ spin state,
\begin{equation}
\ket{\psi} = \textstyle \frac{1}{\sqrt{N_{\mathrm{I}}}} \displaystyle
\left(\ket{\downarrow \downarrow \downarrow \cdots \downarrow} +
\ket{\uparrow \uparrow \uparrow \cdots \uparrow} \right), \label{GHz}
\end{equation}
and the same protocol could be used as for the single ion to accumulate
phase, but with the resolution (and sensitivity) enhanced by a factor
of $N_{\mathrm{I}}$.  These states (\ref{GHz}) have been created for
as many as $N_{\mathrm{I}} = 14$ ions \cite{MonzPRL11}, and multiple
groups are actively pursuing various ways to scale up the size of
entangled trapped ion systems.

\ack We thank Amar Vutha, Chris Monroe, Dana Anderson, and Kale
Johnson for helpful discussions.  W.C.C.  Acknowledges support from
the U.S. Army Research Office under award W911NF-15-1-0261 and
University of California Office of the President's Research Catalyst
Award No. CA-15-327861.  P.H. acknowledges support from the University
of California Office of the President's Research Catalyst Award
No. CA-16-377655.

\section*{References}
\bibliography{IonGyro}

\appendix
\section{Area formula}\label{AreaAppendix}
We show using semiclassical derivation that the area enclosed by the
interferometer is insensitive to the ion's initial position and
momentum in $x$ and initial momentum in $y$. The area for a trajectory
enclosed by a periodic trajectory $\mathbf{r}(t)$ is given by the path
integral
\begin{eqnarray}
\mathbf{A} &=
\frac{1}{2}\oint\mathbf{r}\times\,\mathrm{d}\mathbf{r}=\frac{1}{2}\int_0^T
\mathrm{d}t\,\mathbf{r}(t)\times\mathbf{v}(t)\\
& =\frac{1}{2m}\int_0^T \mathrm{d}t\, \mathbf{J}=\frac{\mathbf{J}\,T}{2m}
\end{eqnarray}
where $T \equiv 2 \pi/\omega$ is the period and $\mathbf{J}$ is the
angular momentum.  A momentum kick, $\Delta\mathbf{p}$, at the start
of the trajectory (and taken to be along $x$) changes the angular momentum by
$\Delta\mathbf{J}_{\mathrm{SDK}} = \mathbf{r}(0)\times\Delta\mathbf{p}
= \mathbf{r}_\perp(0)\times\Delta\mathbf{p}$, where
$\mathbf{r}_\perp(0)$ is the component of the initial displacement
perpendicular to the direction of the momentum kick (which we will
take to be the $y$-direction).  The trap
displacement in $y$ then changes the angular momentum by $\Delta
\mathbf{J}_{\mathrm{d}} = -y_{\mathrm{d}} \mathbf{\hat{y}} \times
\mathbf{p}(0)$. We are interested in the areas for two trajectories
with initial momentum kicks, $\pm \Delta \mathbf{p} =\pm
N_{\mathrm{k}}\hbar\Delta k \,\mathbf{\hat{x}}$, in opposite directions in $x$.  The area
enclosed by the interferometer is the difference between these areas,
taken over half a motional period (since the interferometer closes at
time $T/2$):
\begin{equation}
A =
\left|\frac{\Delta\mathbf{J}\,\frac{T}{2}}{2m}\right|= \pi
\frac{\Delta p}{m \omega} (y_{\mathrm{d}} - r_{\perp} (0) ),
\end{equation}
which agrees with (\ref{Area}).  We see that the area difference depends
only on the initial displacement perpendicular to the SDK direction
and the size of the kick.  The formula holds for circular, elliptical,
or even straight line trajectories and is also independent of the
initial momentum of the particle.

\end{document}